\documentclass[a4paper]{jpconf}
\usepackage{graphicx}
\begin{document}
\title{Fifty Years of Quasars: Physical Insights and Potential for Cosmology}
\def\nat{Nature\/}
\def\apj{ApJ\/}
\def\aap{AAp\/}
\def\mnras{MNRAS}
\def\apjl{ApJL\/}

\author{J. W. Sulentic$^{1}$, P. Marziani$^{2}$, D. Dultzin$^{3}$, M. D'Onofrio$^{4}$,\\ A. del Olmo$^{1}$}

\address{$^{1}$Instituto de Astrof\'{\i}sica de Andaluc\'{\i}a (IAA-CSIC), Spain\\ $^{2}$INAF Oss. Astronomico Padova, Italy\\ $^{3}$IA-UNAM, Mexico\\
$^4$Dipartimento di Fisica \& Astronomia, Universit\`a di Padova, Italy}

\ead{sulentic@iaa.es}

\begin{abstract}
Last year (2013) was more or less the 50th anniversary of the discovery of quasars.
It is an interesting time to review what we know (and don't know) about them both
empirically and theoretically. These compact sources involving line emitting plasma
show extraordinary luminosities extending to one thousand times that of our Milky Way
in emitting volumes of a few solar system diameters (bolometric luminosity $\log L_\mathrm{bol} \sim $ 44 - 48 [erg 
s$^{-1}$]: $D$=1-3 light months $\sim 10^3 -10^4$\ gravitational radii). 
The advent of 8-10 meter class telescopes enables us to study them
spectroscopically in ever greater detail.

In 2000 we introduced a 4D Eigenvector 1 parameters space involving optical,
UV and X-ray measures designed to serve as a 4D equivalent of the 2D 
Hertzsprung-Russell diagram so important for depicting the diversity of stellar types and
evolutionary states. This diagram has revealed a principal  sequence of
quasars distinguished by Eddington ratio (proportional to the accretion rate 
per unit mass).  Thus while stellar differences are primarily driven by the mass 
of a star, quasar differences are apparently driven  by the ratio of luminosity-to-mass.

Out of this work has emerged the concept of two quasars populations A and B
separated at Eddington ratio around 0.2 which maximizes quasar multispectral differences.
The mysterious 8\%\ of quasars that are radio-loud belong to population B which
are the lowest accretors with the largest black hole masses. Finally we
consider the most extreme population A quasars which are the highest
accretors and in some cases are among the youngest quasars. 
We describe how these sources might be exploited as standard candles for cosmology.
\end{abstract}

\section{Introduction}

\begin{figure*}[htp!]
\includegraphics[scale=0.45]{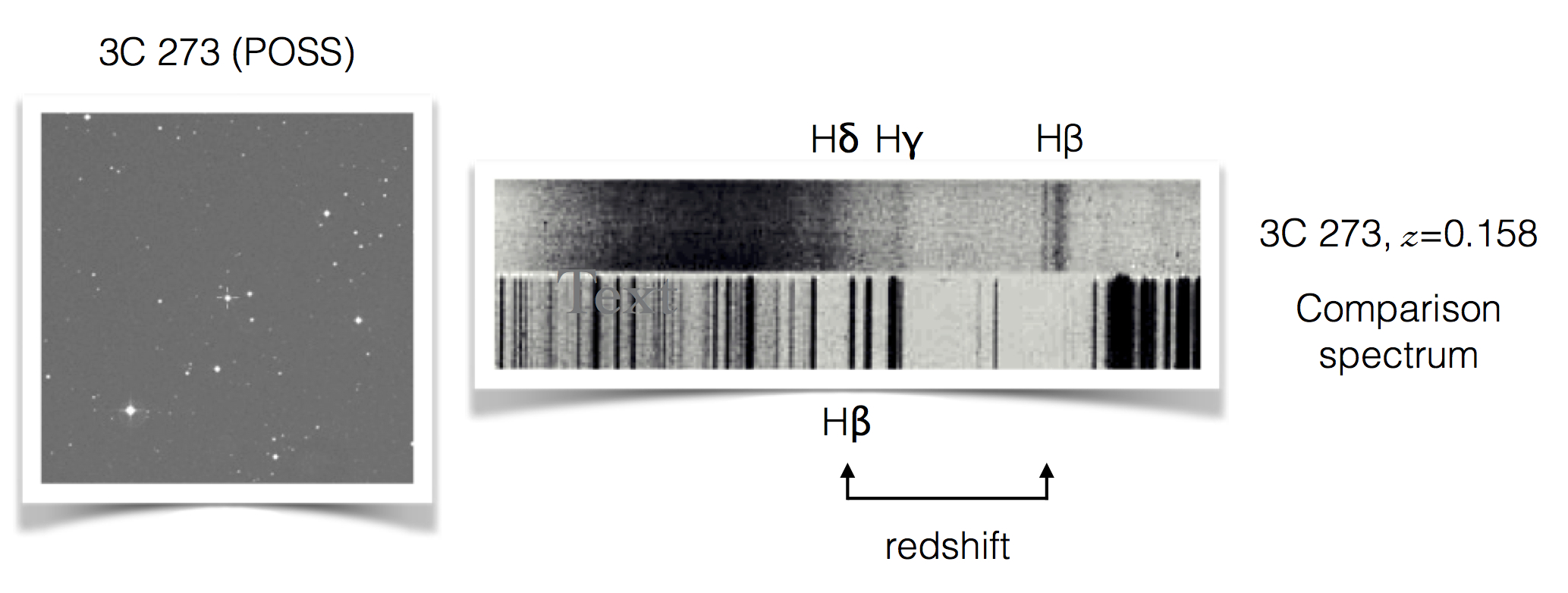}
\caption{\label{fig:3c273} Left: An optical image of 3C273 the first (or one of the first) quasars
discovered. It was radio-loud and also showed an optical jet. Right: optical spectrum of 3C273 
showing broad Permitted and narrow  forbidden emission lines. Optical FeII emission is unusually 
strong in this relatively nearby quasar--a harbinger of the FeII emission now seen in almost all 
low $z$\ quasars. }
\end{figure*}

Today the existence of a large quasar population is taken for granted. They 
are simply regarded as the hyperactive nuclei of galaxies observed across 
cosmic time. In 1963 the reaction was quite different. In the years 
immediately before the discovery, some blue stars  had apparently been
detected with radio telescopes. This was a surprise because stars are such 
weak radio emitters (thermal radio emission from the Sun was only detected 
around 1946 \cite{hey46}. These mysterious radio stars were eventually observed 
spectroscopically and they showed most unusual spectra. Instead of typical stellar 
absorption line spectra they showed broad emission lines \cite{schmidt63,schmidtmatthews64}. 
At first the lines could not be identified but after some time it was realized that 
they were redshifted Balmer lines (in the first two cases $z= c\Delta \lambda/\lambda_0$= 0.16 (3C 273) 
and 0.42 (3C 47)). In the latter case MgII2798 was detected -- it had only been 
detected in the {\it sun} previously. In those early years even the cosmological nature 
of the quasar redshift was an open question although a gravitational origin for the 
redshift had been ruled out \cite{greensteinschmidt64}. If the redshift reflected 
velocity of recession in an expanding Universe then the quasars were extremely 
luminous--and small-- based on emitting region sizes (a few thousand Schwarzschild 
radii) inferred  from timescale of intensity fluctuations (weeks/months). Figure \ref{fig:3c273}
shows an optical image and spectrum for 3C273.

\begin{figure*}[htp!]
\includegraphics[scale=0.45]{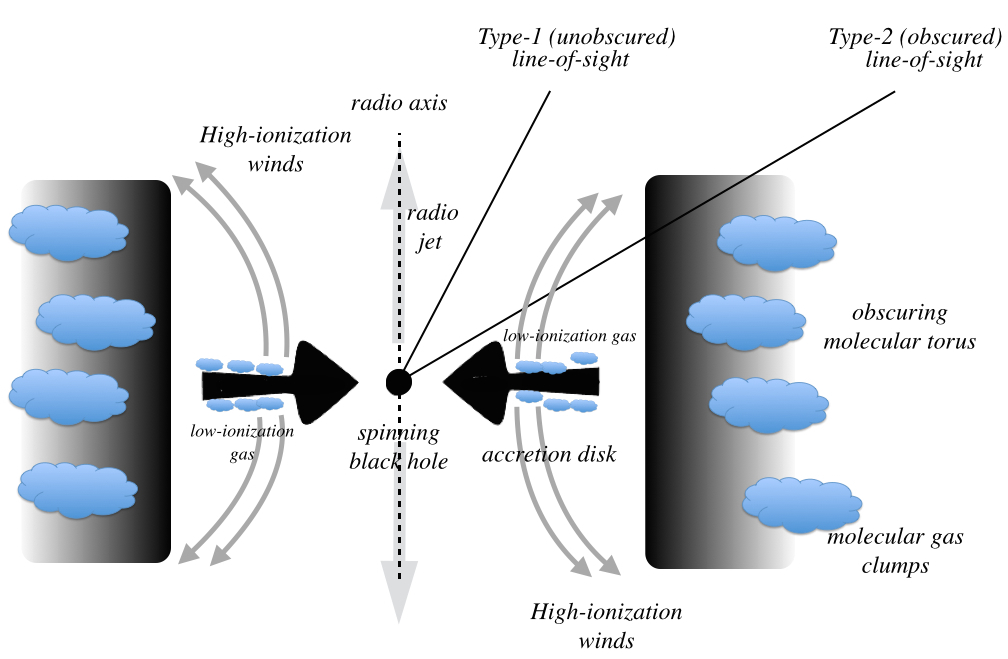}
\caption{\label{fig:cartoon} A cartoon illustrating AGN accretion and unification ideas. 
Broad lines (i.e. quasars) are seen in sources oriented at intermediate viewing angles. 
Narrow emission lines can be seen in all or most sources oriented near edge-on because
the lower density gas extends high enough to be seen above or below the torus. The
torus blocks the broad line emission from the accretion disk.  The disk is shown with  an
``arrow'' shape indicating that there are many models involving inner radiative and outer 
gravitational instabilities for this structure. We show a disk with  an inner geometrically 
and optically thick structure (also known as a slim disk \cite{abramowiczetal88}) 
expected to form only for dimensionless accretion rates $\dot{m} > 0.2  - 0.3$. High ionization 
outflows emerge from the disk, while lower ionization gas may be confined to a flattened 
configuration that is coplanar with the disk (see, e.g., 
\cite{marzianisulentic12a,marzianisulentic12} for reviews and references).  }
\end{figure*}

Ideas about this new constituent of the Universe were in great flux during these 
early years. It is important to remember that quasars were discovered 20 
years before the advent of charge-coupled devices (CCD) detectors and space astronomy was in its infancy. 
Computers and sophisticated electronic instrumentation were yet to come. The 
first quasar analog spectra were recorded on glass plates (a nonlinear detector 
with 1-2\% quantum efficiency). It is difficult from the perspective of 2014 
to imagine that any progress could have been made. Already in 1967 a quasar 
with $z = $2.2 was found (the record is now $z \approx$ 7.08; \cite{mortlocketal11}) showing 
broad and narrow emission lines as well as absorption lines with different redshifts 
\cite{arpetal67}. This is perhaps one of the reasons why the nature of the redshift 
was an open question from the beginning. By about 1970 however a consensus 
had been reached about the nature of such sources -- they were powered by 
gravitational accretion onto a supermassive black hole 
(\cite{zeldovicnovikov64,salpeter64,lyndenbell69}; the model: a line emitting 
accretion disk (AD) surrounding a supermassive black hole (BH): AD+BH, \cite{shakurasunyaev73}). By 1985 an obscuring 
torus had been added to the standard model (\cite{nenkovaetal08} and references therein). 
A unification model involving a 
BH+AD+torus structure viewed from different directions can indeed unite much 
of the active galactic nuclei (AGN) diversity (the unifying name for all the manifestations of quasar 
activity ``active galactic nucleus"  came into use in the early  '80s).  Figure \ref{fig:cartoon} sketches the main
elements of the standard accretion model. Not all elements are in the cartoon
need to be present in the same AGN: the innermost thick structure (the slim disk) is expected to develop only for 
relatively high accretion rates, above Eddington ratio $\approx 0.2 - 0.3$. Some type-2 AGN may be sources accreting at very low rate,   missing broad line emitting gas and not obscured by a thick torus of clumpy molecular gas  \cite{laor03}.

\section{Definition of a quasar}

Physicists sometimes speak about the need for establishing an operational definition of a 
phenomenon in order to facilitate development of models employing the known laws of physics. 
In astronomy, where sources are remote and observations beset with large uncertainties,
theoretical speculations are often far ahead of the evidence. Our 1963-64 definition of quasar 
would involve something like "radio-loud blue variable stellar (i.e. point) source showing broad 
redshifted emission lines" . The first quasars showed a UV excess reflecting 
a continuum rising toward shorter wavelengths which is well visible as the dark band short-ward 
of H$\delta$ on the spectrum of Figure \ref{fig:3c273}. This kind of spectral energy distribution 
is similar to the one observed from "hot" white dwarf stars. Searches were rapidly devised to 
find more very blue stars. Followup spectroscopy could then separate the white dwarf stars 
from the high redshift quasi-stellar sources. This kind of optical selection quickly demonstrated 
that radio-quiet quasars outnumbered radio-loud ones. Today only a mysterious 8\% of quasars  
with $z<$1.0 are found to be radio-loud. ``Radio" cannot be part of an operational definition 
for quasars; neither ``blue" because quasars are now found with a wide range of colors. 
Note that in astronomy what you see is often affected by the intervening medium--all or part 
of the dispersion in a property might not be intrinsic to a source (i.e. the Sun 
appears redder when it is observed at low altitude).

The word "stellar" also becomes a problem for our operational definition 
if we realize that evidence for broad emission lines in the nuclei of some 
local galaxies had been known since early in the 20th century (hence the 
label AGN). Most notable are local spiral galaxies with ultra-luminous nuclei 
\cite{seyfert43}. A conference was held at Steward Observatory in 1968 to debate 
the idea that Seyfert galaxy nuclei might simply be lower luminosity manifestations 
of the quasar phenomenon. Quasars at high redshift appear stellar because their 
host galaxies are too faint to be detected. Hubble Space Telescope has revealed 
evidence of host galaxies around many high redshift quasars. A few claims of naked 
quasars remain intriguing (e.g. \cite{magainetal05}). It turns out that many quasars 
and Seyfert galaxies require us to remove "broad emission" from the operational
definition  as well because many of these sources show only narrow (forbidden) 
lines. Later the discovery of lineless quasars (i.e. BL Lacs) made our original 
definition even more off the mark \cite{schmidt68,strittmatteretal72}. All or 
much of our problems with definition stem from the likelihood that quasars, 
unlike stars, are not isotropic sources. They presents different aspects when 
viewed at different orientations to the line of sight. Today all of the manifestations 
found before, during and after 1963 are united under the label "active galactic 
nuclei" (AGN) a term which came into usage in the early '80s. Much detailed 
history, and a full discussion of nomenclature, can be found in  
"Fifty Years of Quasars" \cite{donofrioetal12}.

X-ray astronomy was born with the launch of UHURU just seven years after quasars 
were discovered (3C273 had already been X-ray detected: \cite{bowyeretal70}).
The best empirical commonality (operational definition?) for AGN may involve 
the presence of a hard X-ray power-law component (e.g. \cite{elvisetal78}).
It is important to point out that the central AD+BH+torus structure
cannot be spatially resolved in any AGN even if interferometric mid-infrared observations are likely to achieve this goal \cite{jaffeetal04}. Even if a few of the lowest redshift examples 
are eventually resolved it is not clear how much this will advance our understanding
of central source structure. All is not lost of course--we can use spectroscopy  
to resolve the structure in many quasars--most are relatively 
unobscured essentially down to the BH event horizon (emission lines are seen at radii 
as small as a few hundred $R_G$ in many sources--and possibly to 2 -- 6  $R_G$  via the
6.4keV Fe K$\alpha$\ line in a few cases).

\section{CONTEXTUALIZING QUASARS -- TOWARDS A QUASAR H-R DIAGRAM?}

Given the multiwavelength diversity of AGN properties now observed it 
should come as no surprise that different groups now focus on specific 
subtypes (e.g. blazars, radio galaxies, LINERS, narrow line Seyfert 1s)
and even on observations in a particular wavelength domain.
It can be educational to use the NASA Extragalactic Database (NED) search 
engine to identify all papers discussing one of the hundred best studied sources. 
In many cases one finds a strong bias towards radio studies (followed by  IR and X-ray)
perhaps because after 50 years we are still unclear if all quasars pass 
through a RL phase or if the 8\%\ of RL quasars have fundamentally different 
source central structure/kinematics diversity 

Another important component missing from quasar studies until the '90s  
(or we would say the 2000s) was an empirical formalism allowing one to 
contextualize quasar  (especially spectroscopic) diversity. When making a 
NED search one often finds only 1-2 out of 100-500 references for a specific 
source presenting/discussing optical spectra. One reason for this bias is 
probably the widespread impression that all quasar spectra are self similar. 
Nothing could be further from the truth. It is our contention that this 
misimpression has seriously retarded progress on quasar studies. 
 
\begin{figure*}[htp!]
\includegraphics[scale=0.6]{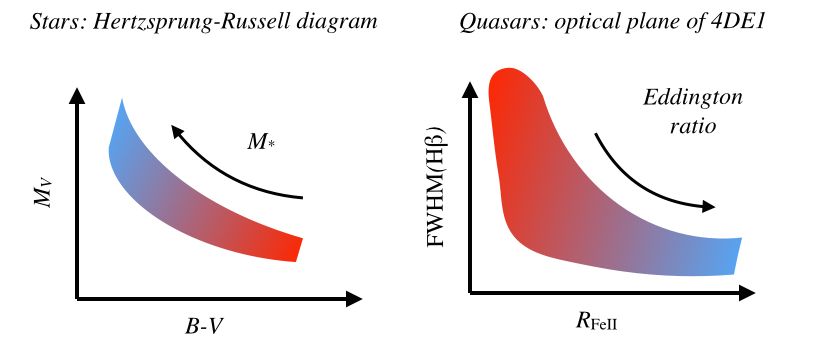}
\caption{\label{fig:hr}  Cartoon representations of the 2D stellar H-R diagram (left)
and 2 of the 4 dimensions (i.e. optical plane) of the 4D Eigenvector 1 (4DE1) diagram
for quasars. In  both cases the schematic main sequence of source occupation is shown
with  the stellar sequence drive by mass (M) and the quasar sequence driven
by Eddington ratio. More dimensions are needed for quasars because, unlike stars, 
they are not isotropic radiators}
\end{figure*}

A good and simple example of source contextualization involves the 
Hertzsprung-Russell diagram for stars (Figure \ref{fig:hr}). If all stellar spectra were self similar such a 
contextualization would not be useful or even possible. But we recognize at 
least seven -- OBAFGKM -- principal stellar spectral types organized along 
a Main Sequence as well as specific regions of the diagram occupied by stars 
in particular states of evolution. The principal driver is recognized to be 
stellar mass with different fusion processes and metallicity playing important 
roles as well. Can the much more energetic quasar phenomenon contain so little 
spectroscopic diversity that contextualization not useful? How has such an impression 
come about? In part it reflects the scarcity of good spectroscopic data 
before 1990. In 1989 only about 60 quasars could be found  in the literature 
with spectra suitable for serious classification but even these revealed 
interesting diversity in line profile shape \cite{sulentic89}. The lack of good 
spectra fed the impression that all quasars are spectroscopically similar 
(perhaps in the face of the striking broad redshifted emission lines all 
the rest seemed minor details). We think that the lack of clarity about 
quasar spectral diversity is responsible for our failure, up to the present, 
to develop a standard physical model for the broad line emitting region of 
quasars (e.g. \cite{dumontetal98}). The right panel of Figure \ref{fig:hr}
shows the principal occupation sequence for quasars in a simple plot 
involving two fundamental optical spectroscopic measures.

Serious progress on the spectroscopic front begins after 1992 with an increase
in the quantity and quality (S/N$\sim$20; R$\sim$1000) of 
quasar spectra thanks mostly to the CCD detector which entered astronomy 
in 1982. Lick Observatory was ahead of the game in the sense that 
their IDS (image dissector scanner) system gave them the competitive edge 
in the 1970s-80s. Most of this early work focussed on the brighter Seyfert 1 nuclei. 
Quasar spectroscopy had to wait for the CCD era (except for the image photon counting system (IPCS) 
used at Palomar in this period). 

From this point onwards let us focus on type 1 AGN (quasars and Seyfert 1 galaxies) 
which show both high and low ionization broad emission lines. They also almost always 
show strong blends of permitted FeII emission--often enough that one might  want 
to keep separate the few sources that do not show it. In this way one part of 
an operational definition for (low redshift) Type 1 quasars can be ``broad redshifted 
low ionization emission lines including optical FeII". The FeII requirement cannot be 
confirmed in sources beyond z=0.7 unless we: 1) accept UV FeII emission as a surrogate 
for optical FeII emission and/or 2) follow the increasingly redshifted region of the 
strongest optical FeII blends near H$\beta$ into the infrared windows. This has been 
accomplished for about 70 quasars in the range z=1.0-3.7 using the VLT-ISAAC infrared 
spectrometer \cite{sulenticetal04,sulenticetal06a,marzianietal09}. FeII emission is seen in 
all of them \cite{marzianietal09} while a study of almost 500 of the brightest low redshift
quasars in the northern sky found all but 14 with detectable FeII emission \cite{zamfiretal10}.

Our focus on Type 1 sources reflects a working strategy that they
are the high accreting {\bf parent population} of the AGN phenomenon.
Inclusion of Seyfert 1 sources with quasars requires an addition to the operational definition 
"stellar sources with and without evidence of a surrounding host galaxy". We 
assume that the broad emission lines arise in a flattened distribution of 
gas associated with an accretion disk (AD). The AD provides a dense 
medium suitable for FeII production. Permitted FeII emission is strong in most type 1 
sources and its strength requires a low ionization, high density,
and high column density  emitting medium. The best candidate for such 
an emitting region is arguably the AD. There is also evidence that part of the Balmer lines arise 
in the clouds emitting Fe II. Gravitational accretion onto 
a supermassive BH is the only know mechanism that can account for the 
extreme luminosities of the Type 1 sources. Hence the BH+AD paradigm.

Restriction to Type 1  sources is required for this type of approach. Following 
everything that we know they represent the most unambiguous class of high 
accreting sources offering the most spectroscopic clues into their nature.
Once they are fully characterized and understood  one can try to unify them with
other more ambiguous classes of AGN with and without broad lines.
A further restriction to sources with broad emission lines AND optical FeII 
emission removes additional ambiguities associated with obscured sources 
(e.g. Seyfert 1.5) and sources with the narrowest broad lines (e.g. narrow line 
Seyfert 1 (NLSy1) sources). Even with these restrictions we find impressive 
spectroscopic diversity. Understanding this diversity --we argue--is a key to 
development of successful physical models for the broad line region (BLR) of Type 1 sources.

Progress in understanding Type 1 quasars came with the advent of larger 
samples of quasars with high s/n spectra. A major advance involved analysis 
of spectra for 87 sources in the  Palomar Bright Quasar survey \cite{borosongreen92}. 
The spectra allowed accurate measurements of broad and narrow emission line 
properties. Principal Component Analysis (PCA) techniques were applied to 
the correlation matrix representing the measured parameters. This study 
identified the principal (Eigenvector 1) correlations that exist 
in the dataset and also showed that source luminosity belonged to 
Eigenvector 2 (the second orthogonal solution). This work (and also 
work on X-ray-optical correlations \cite{wangetal96} motivated us to 
develop the empirical context in which to interpret the spectroscopic 
diversity of type 1 sources \cite{marzianietal96,sulenticetal00a}. Fortunately 
by the mid 90s the Hubble Space Telescope archive was supplying 
moderate quality UV spectra for more than 130 low z sources.

\section{Our work with 4D eigenvector 1 -- major results}

We focussed our efforts on a four dimensional parameter space (4D 
Eigenvector 1 = 4DE1) \cite{sulenticetal00b}.
4DE1 has roots in the PCA analysis of the Bright Quasar Sample
(87 sources, \cite{borosongreen92}, hereafter BG92) as well as in correlations
that emerged from ROSAT(e.g.,\cite{wangetal96}). 4DE1 as we
define it involves BG92 measures: (1) full width half maximum of
broad H$\beta$\ (FWHM H$\beta$) and (2) equivalent width ratio of optical
FeII\ and broad H$\beta$: RFe = W(FeII$\lambda$4570)/W(H$\beta$). We added a
\cite{wangetal96}-defined measure involving (3) the soft X-ray
photon index ($\Gamma_\mathrm{soft}$) and a  measure of (4) CIV$\lambda$1549\ broad line profile
velocity displacement at half maximum ($c(\case{1}{2})$) to arrive
at our 4DE1 space. Other points of departure from BG92 involve our
comparison of radio-quiet (RQ) and radio-loud (RL) sources as well as subordination of BG92
[OIII]\ measures (although see \cite{zamanovetal02,marzianietal03a}).

The principal diagnostic measures defined above  can be interpreted as physical 
measures involving: 1) velocity dispersion of the low ionization broad 
line (HIL) region. It is standard to assume (and there are consistency arguments 
to support it) that these lines arise from a rotating Keplerian disk making 
FWHM (usually H$\beta$ or MgII2798) a virial estimator of BH mass. 2) RFe 
involves the relative strength of  low-ionization lines (LILs) that are thought to arise in the same AD 
structure. This ratio serves as an estimator of electron density (n$_e$) in the 
LIL gas and the high density requirement for its production supports 
the AD as the source of the line emission. 3) $\Gamma_\mathrm{soft}$\ measures
the strength of a (thermal) soft X-ray photon index. This is thought to be
correlated with the accretion rate. 4) The CIV shift parameter measures the 
amplitude of systematic radial motions in the high ionization line (HIL)
gas. It is sensitive to the amplitude and geometry of an AD wind.  

Many other correlates exist especially for quasars above z=1.0.
We focus this summary on low redshift results and the relevant diagnostic measures.
Our 4DE1 parameter space immediately provided several clues about the kinematics 
and physical conditions in the broad line region. Note that the added dimensionality 
of 4DE1 compared to the stellar H-R diagram is required if for no other reason than 
the fact that the multiwavelength properties of a quasar are orientation dependent. 
This is most obvious for RL sources but ample evidence exists that it is also 
true for the RQ majority. Major clues coming out of 4DE1 are summarized here.

\begin{figure*}[h]
\includegraphics[scale=0.4]{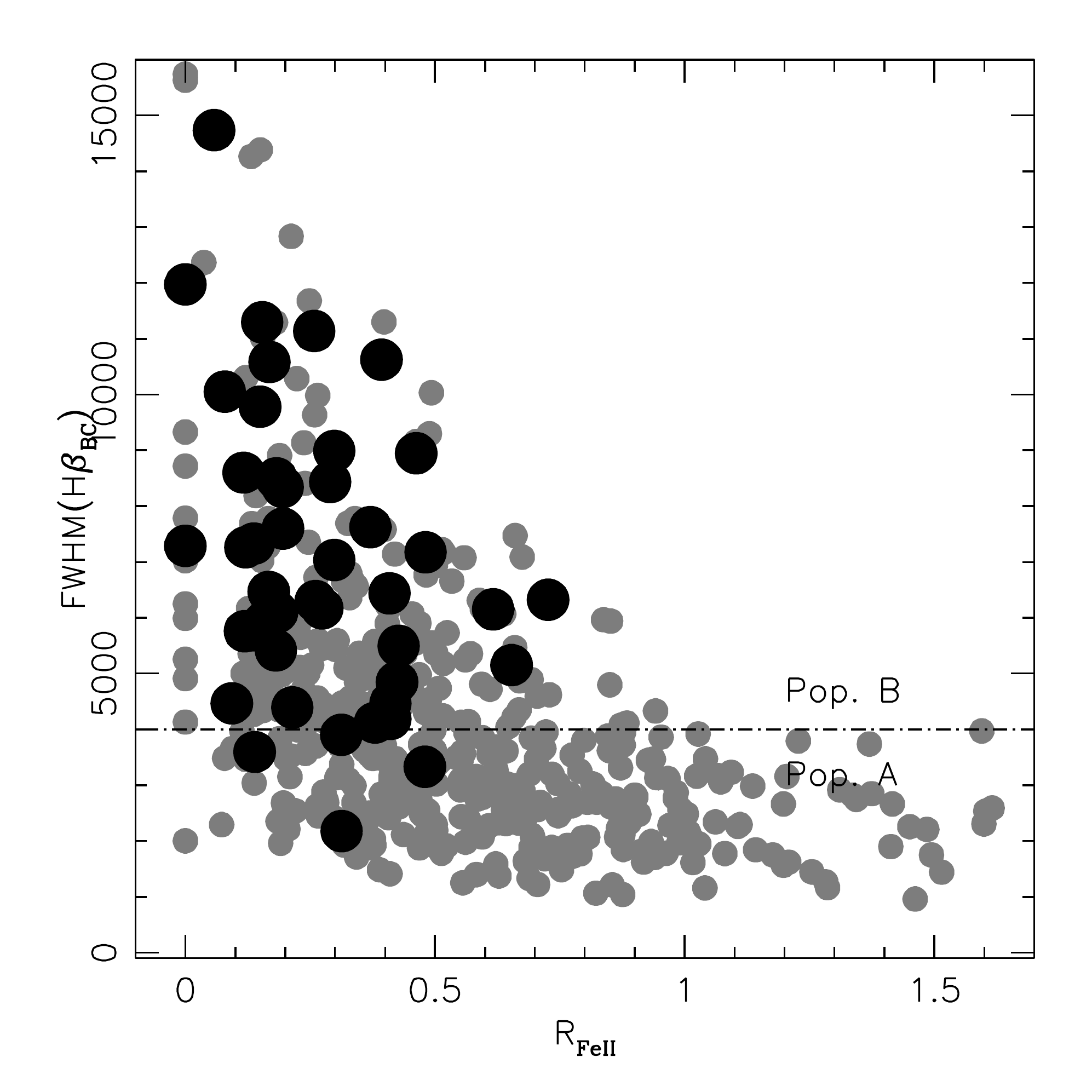}
\includegraphics[scale=0.4]{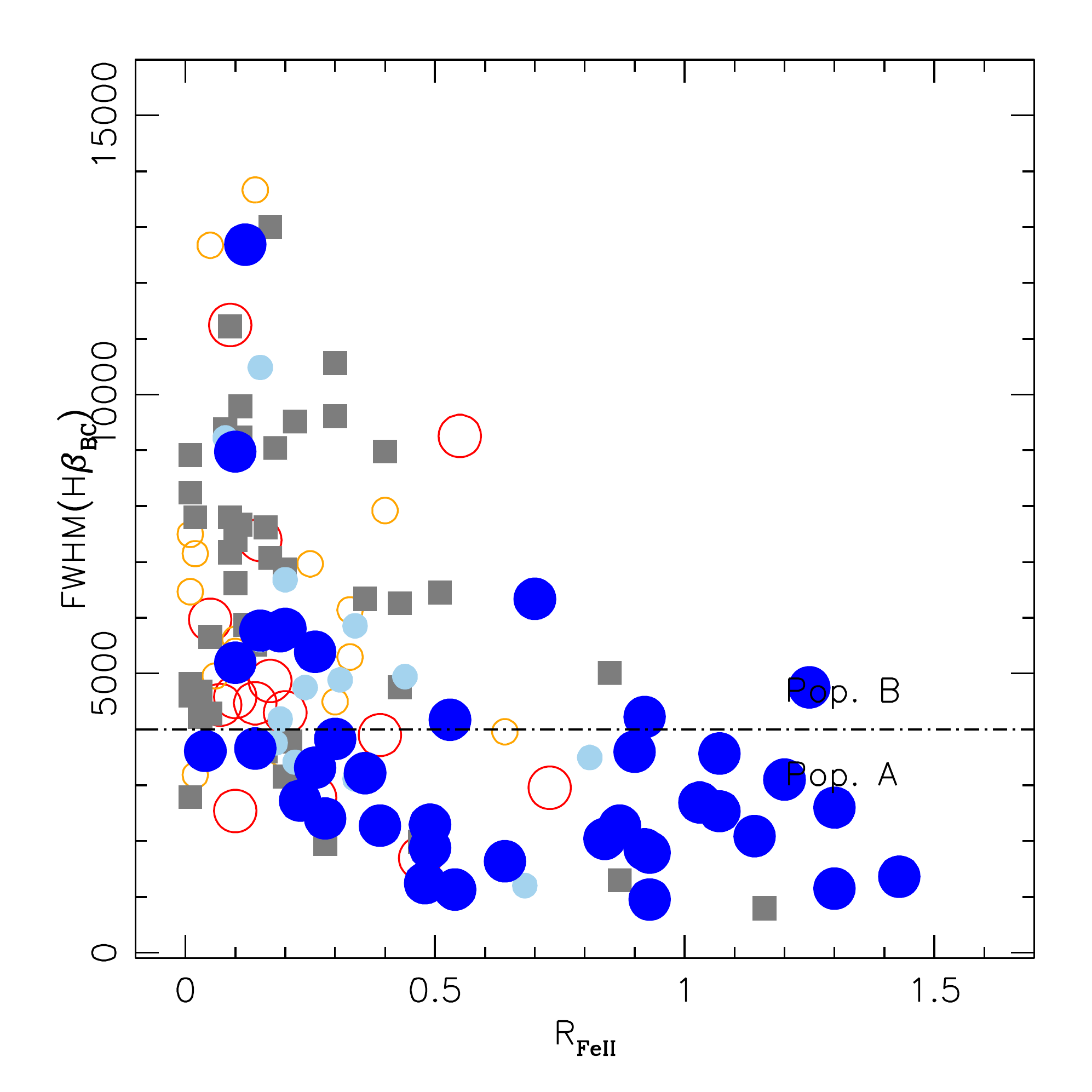}
\caption{\label{fig4}  The 4DE1 optical plane involving the width of H$\beta$ and optical FeII strength. The horizontal line 
marks the boundary between population A (lower) and B (upper) sources. 
Left: source occupation for 470 SDSS-DR5 quasars (z$<$0.75) with highest s/n SDSS spectra \cite{zamfiretal08}. Small grey circles 
represent RQ sources; large black circles lobe-dominated RL  sources which show a strong preference for Population B. 
Right: optical  plane for all low z quasars with measurable HST/FOS UV spectra \cite{sulenticetal07}. Different 
symbols represent the amplitude of the CIV 1549 blueshift at half maximum $c(\frac{1}{2})$. Large blue filled circles
involve sources with the largest CIV blueshifts which strongly favor Population A.  Large open circles represent sources with a 
large CIV redshift and grey squares those with no significant line shift \cite{sulenticetal07,marzianisulentic12}.}
\end{figure*}

\begin{itemize}

\item Figure 4 is the clearest  representation of the diversity of optical and UV spectroscopic measures in low 
redshift sources \cite{sulenticetal00a,sulenticetal00b}. Subsequent work suggests that the trend (i.e. 
source occupation) in Figure 4 (left) is driven by source Eddington ratio (proportional to the dimensionless accretion 
rate for constant radiative efficiency) convolved with line-of-sight orientation. $M_\ast$ (stellar mass) drives the H-R 
diagram main sequence occupation but black hole mass plays a secondary role in 4DE1 \cite{marzianietal01,marzianietal03b} 
in low-$z$ samples. 

Figure 4 (right panel) again shows the 4DE1 optical plane and source occupation for the 
470 SDSS-DR5 quasars with highest s/n SDSS spectra. RL sources are distinguished from 
the RQ majority (grey dots) with symbols that reflect their radio morphology. Lobe 
dominated (LD--largely FRII) sources are indicated by open red circles while 
core-dominated (CD RL) are shown as filled blue squares. The latter are 
interpreted in orientation-unification scenarios as preferentially aligned LD 
where the radio jets are pointed along our line of sight. Our default assumption 
is that the radio jets are aligned perpendicular to the broad line emitting disk.  
The first major result is that the majority of LD sources occupy a restricted domain 
relative to the RQ majority of quasars. The domain is also restricted for the 
CIV and $\Gamma_\mathrm{soft}$\ measures. This may be telling us that RL quasars 
are a distinct quasar population with different BLR structure and kinematics.
A second result involves a displacement of CD RL sources towards smaller FWHM 
H$\beta$ and slightly larger RFE. The green arrow in figure 1 indicates the change 
in median optical 4DE1 measures from LD to CD. This is consistent with the 
orientation-unification ideas mentioned above  and was previously inferred 
from a  correlation between FWHM H$\beta$ and radio core/lobe flux ratio 
\cite{sulenticetal03,zamfiretal08}.  

\item The multiwavelength differences between sources at the two ends of the 
sequence seen in Figure 4 motivated the concept of two quasar populations 
(horizontal line in Figure 4 left/right marks the Population A-B boundary
\cite{zamfiretal08} at FWHM H$\beta$=4000 km s$^{-1}$. This value (and RFE=0.5)  
lie very close to the values identified in a 2D K-S test maximizing the parameter 
space separation between RL and RQ sources. Many other differences exist between 
largely RQ Population A sources (below 4000  km s$^{-1}$) and the mixed RQ/RL population B  
above this FWHM boundary. Physically the FWHM H$\beta$ boundary (4000  km s$^{-1}$) 
corresponds to $L/L_\mathrm{Edd}=0.2 \pm$0.1 for a source with black hole mass log 
$M_\mathrm{BH} 	\sim $ 8.0. This boundary may represent a critical Eddington ratio where the 
BLR undergoes a significant change in structure/kinematics. Even if it has no 
profound physical significance it represents a way to separate quasars into high 
and low accreting samples. The lack of this kind of discrimination 
underlies previously mentioned difficulties with attempts at developing a model 
for BLR physics.

\item The 4DE1 context reveals that the much discussed narrow line Seyfert 1 
(NLSy1) sources are not a distinct class of quasars with unusual properties.
Rather they are  simply the most extreme (i.e. narrowest FWHM H$\beta$)
Population A sources. All evidence points toward  4000 km s$^{-1}$ (and not
2000  km s$^{-1}$) as the more significant boundary for sources with $\log L \le 46$ [erg s$^{-1}$].
The limit is expected to increase for more luminous sources because of the weak, but 
significant, effect of increasing BH mass \cite{dultzinetal11,negreteetal12}. 

\item Other evidence that Population A and B quasars differ includes a fundamental 
change in broad line profile shape near the Population A-B boundary \cite{sulenticetal00a,sulenticetal02,sulenticetal07}.    
The broad H$\beta$ (characteristic low ionization line at low z) profile 
is well fit by a symmetric Lorentzian function in most Pop. A sources. 
Population B sources require a double Gaussian fit including broad 
(FWHM$\sim$ 5000 km s$^{-1}$) relatively unshifted component plus a very-broad 
(FWHM$\sim$10000 km s$^{-1}$) and redshifted (a few thousand  km s$^{-1}$)
component.  CIV1549 (characteristic high ionization line at most 
redshifts) shows strong profile blueshifts and blue asymmetries in 
Population A sources but is usually relatively symmetric and unshifted 
in Population B sources. This description is confirmed in median composites 
involving 50-250 low z sources despite the fact that individual sources show 
an impressive diversity in profiles shapes/shifts/asymmetries.  

\end{itemize}

We think that the above results  offer the key to a physical model of quasars
that can unify the intriguing 4DE1 sequence and the Population A-B differences. 
A major advance in our understanding of quasar physics is coming soon.

\section{A promising future}

The Eddington ratio of a quasar is proportional to the ratio of source luminosity to BH mass. 
If Eddington ratio  and BH mass can be derived from some distance-independent measure it
would be  possible to derive distance-independent quasar luminosities. Quasars are now easily 
detected out to $z = 4$ and those in the range $1 < z < 3 - 4$ are of greatest interest because 
the effect of the cosmic matter density is believed to dominate over the repulsive effect of the 
cosmological constant in this range.   Quasars radiating close to the Eddington limit show distinct 
optical and UV spectral properties that can be recognized in major survey spectra -- 
if the data is contextualized within the 4DE1 formalism. Measures of the H$\beta$ spectral 
range  and the 1900 blend  yield  selection criteria involving two related ratios: 
(i) Al III $\lambda$1860/Si III] $\lambda$1892 $\ge$ 0.5 and (ii) Si III] 
$\lambda$1892/C III]$\lambda$1909 $\ge$1.0. A little less than 100 sources  satisfying these criteria 
cluster around the Eddington limit with a relatively small dispersion $\approx$ 0.13 dex \cite{marzianisulentic14}. 
Other 4DE1 correlated parameters related to the X-ray continuum shape have been proposed \cite{wangetal13,lafrancaetal14} and may become  useful with new data provided by 
planned space missions such as Athena or ongoing ones like nuSTAR. More than 300,000 quasars are catalogued, including the 160,000 whose spectra have been collected in the recent SDSS IV - BOSS survey \cite{parisetal14}. The optical and UV selection criteria offer  the non-negligible advantage of facilitating  selection of large samples from presently available data since  the frequency of quasars radiating close to L/L$_\mathrm{Edd}$=1  is estimated 
to be at least a few percent of the unobscured population. We have shown that   such quasar samples  can yield 
independent measures of $\Omega_\mathrm{M}$\ with tight limits even if samples of just a few hundred 
sources are considered  \cite{marzianisulentic13,marzianisulentic14}.
 
Current issues go beyond the existence of the dark energy and focus more on its properties. The simplest 
model for dark energy is a cosmological constant with a fixed equation of state ($p = w \rho$, with 
fixed $w=-1$). However, the dark energy density may depend weakly on time, according to many proposed 
models of its nature \cite{carroll05}: a general scalar field predicts $w$ to be negative and evolving with 
redshift. The strength of a quasar sample also includes the possibility to construct a Hubble diagram 
uniformly sampling a broad range of redshift. Extreme Eddington quasars are, at least in principle,
sources suitable for testing whether the dark energy equation of state is constant or 
is evolving as function of redshift following selected parametric forms for $w(z)$). 
After 50 years, and a few failed attempts, are we on the threshold of using quasars for cosmology?


\section*{References}
\bibliographystyle{iopart-num}

\begin{thebibliography}{10}
\expandafter\ifx\csname url\endcsname\relax
  \def\url#1{{\tt #1}}\fi
\expandafter\ifx\csname urlprefix\endcsname\relax\def\urlprefix{URL }\fi
\providecommand{\eprint}[2][]{\url{#2}}

\bibitem{hey46}
{Hey} J~S 1946 {\em \nat\/} {\bf 157} 47--48

\bibitem{schmidt63}
{Schmidt} M 1963 {\em Nature\/} {\bf 197} 1040

\bibitem{schmidtmatthews64}
{Schmidt} M and {Matthews} T~A 1964 {\em \apj\/} {\bf 139} 781--785

\bibitem{greensteinschmidt64}
{Greenstein} J~L and {Schmidt} M 1964 {\em ApJ\/} {\bf 140} 1--34

\bibitem{mortlocketal11}
{Mortlock} D~J, {Warren} S~J, {Venemans} B~P, {Patel} M, {Hewett} P~C,
  {McMahon} R~G, {Simpson} C, {Theuns} T, {Gonz{\'a}les-Solares} E~A, {Adamson}
  A, {Dye} S, {Hambly} N~C, {Hirst} P, {Irwin} M~J, {Kuiper} E, {Lawrence} A
  and {R{\"o}ttgering} H~J~A 2011 {\em \nat\/} {\bf 474} 616--619
  (\textit{Preprint} \eprint{1106.6088})

\bibitem{arpetal67}
{Arp} H~C, {Bolton} J~G and {Kinman} T~D 1967 {\em \apj\/} {\bf 147} 840--845

\bibitem{zeldovicnovikov64}
{Zel'dovich} Y~B and {Novikov} I~D 1964 {\em Soviet Physics Doklady\/} {\bf 9}
  246

\bibitem{salpeter64}
{Salpeter} E~E 1964 {\em \apj\/} {\bf 140} 796--800

\bibitem{lyndenbell69}
{Lynden-Bell} D 1969 {\em \nat\/} {\bf 223} 690--694

\bibitem{shakurasunyaev73}
{Shakura} N~I and {Sunyaev} R~A 1973 {\em \aap\/} {\bf 24} 337--355

\bibitem{nenkovaetal08}
{Nenkova} M, {Sirocky} M~M, {Nikutta} R, {Ivezi{\'c}} {\v Z} and {Elitzur} M
  2008 {\em \apj\/} {\bf 685} 160--180 (\textit{Preprint} \eprint{0806.0512})

\bibitem{laor03}
{Laor} A 2003 {\em \apj\/} {\bf 590} 86--94


\bibitem{seyfert43}
{Seyfert} C~K 1943 {\em \apj\/} {\bf 97} 28--40

\bibitem{magainetal05}
{Magain} P, {Letawe} G, {Courbin} F, {Jablonka} P, {Jahnke} K, {Meylan} G and
  {Wisotzki} L 2005 {\em \nat\/} {\bf 437} 381--384 (\textit{Preprint}
  \eprint{arXiv:astro-ph/0509433})

\bibitem{schmidt68}
{Schmidt} M 1968 {\em \apj\/} {\bf 151} 393--410

\bibitem{strittmatteretal72}
{Strittmatter} P~A, {Serkowski} K, {Carswell} R, {Stein} W~A, {Merrill} K~M and
  {Burbidge} E~M 1972 {\em \apjl\/} {\bf 175} L7--L13

\bibitem{donofrioetal12}
{D'Onofrio} M, {Marziani} P and { Sulentic} J~W (eds) 2012 {\em Fifty Years of
  Quasars From Early Observations and Ideas to Future Research\/} ({\em
  Astrophysics and Space Science Library\/} vol 386) (Berlin-Heidelberg:
  Springer Verlag)

\bibitem{bowyeretal70}
{Bowyer} C~S, {Lampton} M, {Mack} J and {de Mendonca} F 1970 {\em \apjl\/} {\bf
  161} L1--L7

\bibitem{elvisetal78}
{Elvis} M, {Maccacaro} T, {Wilson} A~S, {Ward} M~J, {Penston} M~V, {Fosbury}
  R~A~E and {Perola} G~C 1978 {\em \mnras\/} {\bf 183} 129--157

\bibitem{jaffeetal04}
{Jaffe} W, {Meisenheimer} K, {R{\"o}ttgering} H~J~A, {Leinert} C, {Richichi} A,
  {Chesneau} O, {Fraix-Burnet} D, {Glazenborg-Kluttig} A, {Granato} G~L,
  {Graser} U, {Heijligers} B, {K{\"o}hler} R, {Malbet} F, {Miley} G~K,
  {Paresce} F, {Pel} J~W, {Perrin} G, {Przygodda} F, {Schoeller} M, {Sol} H,
  {Waters} L~B~F~M, {Weigelt} G, {Woillez} J and {de Zeeuw} P~T 2004 {\em
  \nat\/} {\bf 429} 47--49

\bibitem{abramowiczetal88}
{Abramowicz} M~A, {Czerny} B, {Lasota} J~P and {Szuszkiewicz} E 1988 {\em
  \apj\/} {\bf 332} 646--658

\bibitem{marzianisulentic12a}
{Marziani} P and {Sulentic} J~W 2012 {\em The Astronomical Review\/} {\bf 4}
  04, 33--57 (\textit{Preprint} \eprint{1210.2059})

\bibitem{marzianisulentic12}
{Marziani} P and {Sulentic} J~W 2012 {\em NARev\/} {\bf 56} 49--63
  (\textit{Preprint} \eprint{1108.5102})

\bibitem{sulentic89}
{Sulentic} J~W 1989 {\em \apj\/} {\bf 343} 54--65

\bibitem{dumontetal98}
{Dumont} A~M, {Collin-Souffrin} S and {Nazarova} L 1998 {\em \aap\/} {\bf 331}
  11--33

\bibitem{sulenticetal04}
{Sulentic} J~W, {Stirpe} G~M, {Marziani} P, {Zamanov} R, {Calvani} M and
  {Braito} V 2004 {\em A\&Ap\/} {\bf 423} 121--132 (\textit{Preprint}
  \eprint{arXiv:astro-ph/0405279})

\bibitem{sulenticetal06a}
{Sulentic} J~W, {Dultzin-Hacyan} D, {Marziani} P, {Bongardo} C, {Braito} V,
  {Calvani} M and {Zamanov} R 2006 {\em Revista Mexicana de Astronomia y
  Astrofisica\/} {\bf 42} 23--39 (\textit{Preprint}
  \eprint{arXiv:astro-ph/0511230})

\bibitem{marzianietal09}
{Marziani} P, {Sulentic} J~W, {Stirpe} G~M, {Zamfir} S and {Calvani} M 2009
  {\em A\&Ap\/} {\bf 495} 83--112 (\textit{Preprint} \eprint{0812.0251})

\bibitem{zamfiretal10}
{Zamfir} S, {Sulentic} J~W, {Marziani} P and {Dultzin} D 2010 {\em \mnras\/}
  {\bf 403} 1759 (\textit{Preprint} \eprint{0912.4306})

\bibitem{borosongreen92}
{Boroson} T~A and {Green} R~F 1992 {\em ApJS\/} {\bf 80} 109--135

\bibitem{wangetal96}
{Wang} T, {Brinkmann} W and {Bergeron} J 1996 {\em A\&Ap\/} {\bf 309} 81--96

\bibitem{marzianietal96}
{Marziani} P, {Sulentic} J~W, {Dultzin-Hacyan} D, {Calvani} M and {Moles} M
  1996 {\em ApJS\/} {\bf 104} 37--70

\bibitem{sulenticetal00a}
{Sulentic} J~W, {Marziani} P and {Dultzin-Hacyan} D 2000 {\em ARA\&A\/} {\bf
  38} 521--571

\bibitem{sulenticetal00b}
{Sulentic} J~W, {Marziani} P, {Zwitter} T, {Dultzin-Hacyan} D and {Calvani} M
  2000 {\em ApJL\/} {\bf 545} L15--L18 (\textit{Preprint}
  \eprint{arXiv:astro-ph/0009326})

\bibitem{zamanovetal02}
{Zamanov} R, {Marziani} P, {Sulentic} J~W, {Calvani} M, {Dultzin-Hacyan} D and
  {Bachev} R 2002 {\em ApJL\/} {\bf 576} L9--L13 (\textit{Preprint}
  \eprint{arXiv:astro-ph/0207387})

\bibitem{marzianietal03a}
{Marziani} P, {Sulentic} J~W, {Zamanov} R, {Calvani} M, {Dultzin-Hacyan} D,
  {Bachev} R and {Zwitter} T 2003 {\em ApJS\/} {\bf 145} 199--211

\bibitem{zamfiretal08}
{Zamfir} S, {Sulentic} J~W and {Marziani} P 2008 {\em MNRAS\/} {\bf 387}
  856--870 (\textit{Preprint} \eprint{0804.0788})

\bibitem{sulenticetal07}
{Sulentic} J~W, {Bachev} R, {Marziani} P, {Negrete} C~A and {Dultzin} D 2007
  {\em ApJ\/} {\bf 666} 757--777 (\textit{Preprint} \eprint{0705.1895})

\bibitem{marzianietal01}
{Marziani} P, {Sulentic} J~W, {Zwitter} T, {Dultzin-Hacyan} D and {Calvani} M
  2001 {\em ApJ\/} {\bf 558} 553--560 (\textit{Preprint}
  \eprint{arXiv:astro-ph/0105343})

\bibitem{marzianietal03b}
{Marziani} P, {Zamanov} R~K, {Sulentic} J~W and {Calvani} M 2003 {\em MNRAS\/}
  {\bf 345} 1133--1144 (\textit{Preprint} \eprint{arXiv:astro-ph/0307367})

\bibitem{sulenticetal03}
{Sulentic} J~W, {Zamfir} S, {Marziani} P, {Bachev} R, {Calvani} M and
  {Dultzin-Hacyan} D 2003 {\em ApJL\/} {\bf 597} L17--L20 (\textit{Preprint}
  \eprint{arXiv:astro-ph/0309469})

\bibitem{dultzinetal11}
{Dultzin} D, {Martinez} M~L, {Marziani} P, {Sulentic} J~W and {Negrete} A 2011
  {Narrow-Line Seyfert 1s: a luminosity dependent definition} {\em Proceedings
  of the conference ''Narrow-Line Seyfert 1 Galaxies and their place in the
  Universe''. April 4-6, 2011. Milano, Italy.\/} Proceedings of Science.
  
\bibitem{negreteetal12}
{Negrete} A, {Dultzin} D, {Marziani} P and {Sulentic} J 2012 {\em ApJ\/} {\bf
  757} id.62, 15 pp. (\textit{Preprint} \eprint{1107.3188})

\bibitem{sulenticetal02}
{Sulentic} J~W, {Marziani} P, {Zamanov} R, {Bachev} R, {Calvani} M and
  {Dultzin-Hacyan} D 2002 {\em ApJL\/} {\bf 566} L71--L75 (\textit{Preprint}
  \eprint{arXiv:astro-ph/0201362})

\bibitem{marzianisulentic14}
{Marziani} P and {Sulentic} J~W 2014 {\em \mnras\/} {\bf 442} 1211--1229
  (\textit{Preprint} \eprint{1405.2727})

\bibitem{wangetal13}
{Wang} J~M, {Du} P, {Valls-Gabaud} D, {Hu} C and {Netzer} H 2013 {\em Physical
  Review Letters\/} {\bf 110} 081301 (\textit{Preprint} \eprint{1301.4225})

\bibitem{lafrancaetal14}
{La Franca} F, {Bianchi} S, {Ponti} G, {Branchini} E and {Matt} G 2014 {\em
  \apjl\/} {\bf 787} L12, 6 pp. (\textit{Preprint} \eprint{1404.2607})


\bibitem{parisetal14}
{P{\^a}ris} I, {Petitjean} P, {Aubourg} E, {Bailey} S, {Ross} N~P, {Myers} A~D,
  {Streblyanska} A, {Bailey} S, {Hall} P B,   {et al.} C 2014 {\em
  AAp} {\bf 563} A54 {\em ArXiv e-prints\/} (\textit{Preprint}
  \eprint{1311.4870})

\bibitem{marzianisulentic13}
{Marziani} P and {Sulentic} J~W 2014 {\em Adv. Space Research
\/} {\bf 54}, 1331--1340  {\em ArXiv e-prints\/} (\textit{Preprint}
  \eprint{13\-10.\-31\-43})

\bibitem{carroll05}
{Carroll} S~M 2005 {\em Observing Dark
  Energy\/} ({\em Astronomical Society of the Pacific Conference Series\/} {\bf
  339} 4--29)  ed {Wolff} S~C and {Lauer} T~R 

\end{thebibliography}
\providecommand{\newblock}{}


\end{document}